# Tripling of the scattering vector range of X-ray reflectivity on liquid surfaces using a double crystal deflector.

*Oleg Konovalov,[1*] Valentina Belova,[1,2] Mehdi Saedi,[3] Irene Groot,[3] Gilles Renaud,[2] Maciej Jankowski[1*]*

[1] *The European Synchrotron- ESRF, 71 Avenue des Martyrs, CS 40220, 38043 Grenoble Cedex 9, France*

[2] *Univ. Grenoble Alpes, CEA, IRIG/MEM/NRS 38000 Grenoble, France*

[3] *Leiden Institute of Chemistry, Leiden University, P.O. Box 9502, 2300 RA Leiden, The Netherlands*

Correspondence e-mail: oleg.konovalov@esrf.fr, maciej.jankowski@esrf.fr

**Abstract:**

We achieved a tripling of the maximum range of perpendicular momentum transfer ($q_z$) of X-ray scattering from liquid surfaces using a double crystal deflector setup to tilt the incident X-ray beam. This is obtained by using Miller indices of the reflecting crystal atomic planes that are three times higher than usual. We calculate the deviation from the exact Bragg angle condition induced by a misalignment between the X-ray beam axis and the main rotation axis of the double crystal deflector and deduce a fast and straightforward procedure to align them. We show measurements of X-ray reflectivity up to $q_z = 7 A^{-1}$ on the bare surface of liquid copper.

**Introduction:**

Investigation of processes occurring at atomic and molecular levels at the surfaces and interfaces of liquids is of paramount importance for fundamental surface science and practical applications in physics, chemistry, and biology (Pershan, 2014; Dong *et al.*, 2018; Zuraiqi *et al.*, 2020; He *et al.*, 2021; Allioux *et al.*, 2022). However, experimental methods that allow insight into these phenomena are scarce, making synchrotron-based X-ray scattering the prime choice when sub-nanometer accuracy is needed. The high intensity of synchrotron X-ray beams, their highly compact beam size, and their very low divergence allow *in situ* and *operando* experiments with sub-second time resolution, which is impossible with standard laboratory X-ray sources. Furthermore, the recent upgrade of the European Synchrotron Radiation Facility (ESRF) allows for very demanding experiments using the extremely bright X-ray source (EBS) with unprecedented parameters (Raimondi, 2016).

One of the most widely used X-ray-based techniques for the characterization of liquid surfaces is X-ray reflectivity (XRR). It relies on measurements of the intensity of the reflected X-ray beam from a surface at varying incidence angles, the so-called reflectivity curve, which is used to deduce the surface's out-of-plane electron density profile. Applications of this method are very diverse. They range from the determination of the roughness of the water surface (Braslau *et al.*, 1985), lipid layers on the water-air interface (Helm *et al.*, 1987), free liquid metal surfaces (Magnussen *et al.*, 1995; Regan *et al.*, 1995) displaying layering, polymer assemblies on water (Kago *et al.*, 1998), to protein layers on liquid surfaces (Gidalevitz *et al.*, 1999). Recent technical developments of advanced sample environments and methods further allowed the investigations of even more complex systems. Among these, we may cite Langmuir troughs (Yun & Bloch, 1989) and specialized reactors (Saedi *et al.*, 2020), studies of electrochemical systems (Duval *et al.*, 2012), layer-by-layer assembly of DNA (Erokhina *et al.*, 2008), self-assembled layers (Bronstein *et al.*, 2022), or 2D materials formation on liquid metal catalysts (Jankowski *et al.*, 2021; Konovalov *et al.*, 2022). Thus, the use of XRR, sometimes in connection with other methods like grazing-incidence small-angle scattering (GISAXS) (Geuchies *et al.*, 2016) or X-ray absorption spectroscopy (XAS) (Konovalov *et al.*, 2020), offers a powerful tool for the characterization of a vast family of materials on liquid surfaces.

However, one general difficulty exists in performing XRR on liquid surfaces since neither the liquid sample nor the synchrotron source can be tilted. The requirement of variation of the X-ray beam grazing angle ($\mu$) at the sample surface to change the (vertical) scattering vector perpendicular to the surface, $q_z = 4\pi\lambda^{-1} \sin\mu$ ($\lambda$ is the X-ray wavelength), brings significant experimental difficulties. Different technical solutions were implemented to overcome this problem. The synchrotron X-ray beam can be inclined with respect to the horizontal sample plane using mirrors or single or double Bragg reflections from crystals (overview in (Pershan & Schlossman, 2012), Chapter 2). The main drawback of using a mirror is the maximum achievable $q_z$ value, usually limited to several critical angles of the total surface reflection on the mirror material. The single crystal deflector (SCD) extends this range to $\mu_{max} = 2\theta$, where $\theta$ is the Bragg angle of the chosen scattering planes of the crystal (Smilgies *et al.*, 2005). However, the use of an SCD demands to move the sample to follow the horizontal and vertical displacement of the beam on it, concomitantly with the change of the $\mu$ angle. This has the drawback to agitate the liquid surface. A more recent solution, the double crystal deflector (DCD) (Honkimäki *et al.*, 2006), relies on a double Bragg reflection from two crystals in a geometry that does not require any sample movement with a change of the $\mu$ angle, thus assuring a more stable measurement. The maximum obtainable incident grazing angle is $\mu_{max} = 2(\theta_2 - \theta_1)$, where $\theta_1$ and $\theta_2$ are the Bragg angles of the first and second crystals, respectively, and $\theta_2 > \theta_1$ (Murphy *et al.*, 2014). Practically, in the case of SCDs or DCDs, the maximum achievable perpendicular momentum transfer $q_z^{max}$, does not depend on the X-ray beam energy (see SI Note 1). The most typical choices of crystal sets used in realized DCDs are Ge(111)/Ge(220), Si(111)/Si(220), and InSb(111)/InSb(220). The maximum scattering vector reached for these sets is about 2.5 Å$^{-1}$ (Honkimäki *et al.*, 2006; Arnold *et*

*al.*, 2012; Murphy *et al.*, 2014), which might not be sufficient for studies of some liquid metals, e.g., the surface layering peak and the first structure peak of liquid copper are present at approximately 3 Å$^{-1}$ (Eder *et al.*, 1980).

The ID10 beamline at ESRF was equipped with an SCD since 1999 (Smilgies *et al.*, 2005). During more than 1.5 decades of operating this instrument, deep technological knowledge and experience were acquired, which led to the design and construction, in collaboration with Huber Diffraktionstechnik GmbH & Co. KG company, of a new generation instrument to study liquid surfaces and interfaces, using a DCD. The new 6+2 diffractometer, equipped with a DCD, has been operating since 2016. This diffractometer has the necessary set of rotation and translation stages to precisely align the DCD and assure its high rigidity and accuracy during operation. In this paper, we present a method of tripling the $q_z^{max}$ value using a DCD by using higher-order Bragg reflections of the two crystals. In practice, we use the Ge(333)/Ge(660) reflections instead of the now standard set of Ge(111)/Ge(220) reflections. In addition, we confirm experimentally that even with a three orders of magnitude loss of photon flux with these reflections, recording X-ray scattering at high $q_z$ is still feasible thanks to the recently upgraded ESRF-EBS synchrotron beam (Raimondi, 2016).

**Experimental:**

XRR measurements using a DCD at the ESRF beamline ID10 were performed using a monochromatic X-ray beam with an energy of 22 keV, monochromatized by Si(111) channel-cut monochromator diffracting in the vertical plane. The DCD was aligned according to the below-described procedure. The beam intensity reaching the sample after scattering by the Ge(333) and Ge(660) reflections was 7·10$^{10}$ ph/s at a synchrotron storage ring current of 200 mA. The full width at half maximum of the beam at the sample position was measured to be 26×10 $\mu$m$^2$ (H×V) after focusing with 29 Be parabolic lenses with a radius of 300 $\mu$m, located before the DCD at 8.9 m from the sample and 36.2 m from the X-ray source. The X-ray beam reflected from the surface was measured with a CdTe MaxiPix 2D photon-counting pixel detector (pixel size: 55×55 $\mu$m$^2$, detector area: 28.4×28.4 mm$^2$, sensor: 1 mm thick CdTe) and 5 s counting time at each incident angle.

We performed XRR measurements on bare liquid copper and on a graphene layer on liquid copper *in situ* at T = 1400 K (above the copper melting temperature) in a specially designed reactor dedicated to chemical vapor deposition (CVD) growth of thin layers of graphene on a liquid metal catalyst (Saedi *et al.*, 2020). Single-layer graphene was grown under the same conditions as described in (Jankowski *et al.*, 2021). The obtained scattering data, which include non-specular components (diffuse scattering and scattering from the bulk of liquid copper), were processed following the procedure presented in

(Konovalov *et al.*, 2022), considering the spread of the beam reflected on the curved surface of the liquid metal.

**Results and discussion:**

The X-ray diffractometer of ID10 is a multi-function device that allows working with bulk and surfaces of solid and liquid samples using different setup geometries, see Fig. 1A and 1B. The X-ray detectors are mounted on the γ and δ circles (see Fig. 1C), allowing their movements around the diffractometer center in horizontal and vertical planes. The available beamline detectors are MaxiPix 2x2 CdTe, Dectris Eiger 4M CdTe, Pilatus 300k Si, Mythen 1K, and Mythen2 2K. The detector holder's construction allows the simultaneous use of these detectors in different configurations during an experiment. The diffractometer comprises two sample stages in horizontal or vertical geometry configuration, see Fig. 1B. The horizontal stage is typically used for the investigation of liquid sample surfaces and comprises three circles θ, χ, and ϕ, and a z-, x-, and y- sample translation stage, marked in Fig. 1C. Similarly, the vertical stage is mounted on the θ circle and comprises three circles ω, χ', ϕ', and a z-, x-, and y- sample translation stage. The diffractometer can be used in two modes. In the first mode, the beam is fixed on the instrument's optical axis, while in the second, the DCD is used to tilt the incoming X-ray beam around the sample plane, see Fig. 1C. The first mode is routinely used to measure solid samples and when the use of a bulky and heavy sample environment is needed, whereas the DCD is used for investigations of liquid surfaces and interfaces.

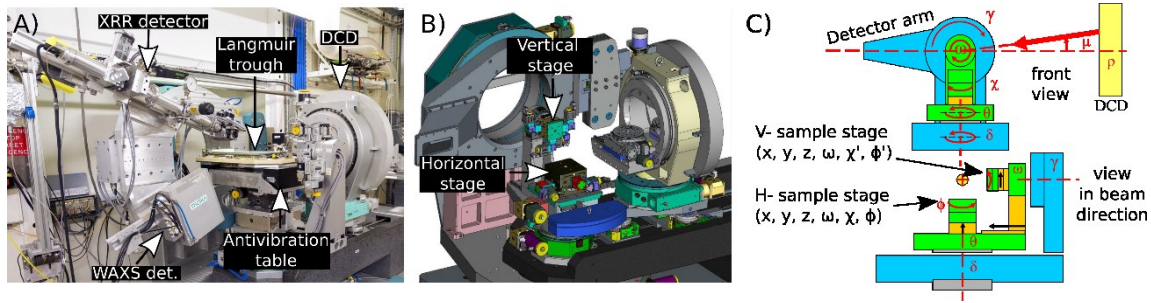

Figure 1: A) Photo of the diffractometer with the mounted Langmuir trough on the antivibration table. Two detectors mounted on the diffractometer arm allow XRR and GIWAXS/GID experiments. B) The 3D drawing of the diffractometer with labeled horizontal and vertical stages. C) Schematic representation of the configuration of the diffractometer circles.

The principle of DCD operation (Honkimäki *et al.*, 2006; Arnold *et al.*, 2012; Murphy *et al.*, 2014) is illustrated in Fig. 2A. The primary incident X-ray beam undergoes a double Bragg reflection by hitting two crystals at points C1 and C2, and at fixed angles θ$_1$ and θ$_2$, respectively, under two constraints. The first constraint imposes that the second Bragg angle is bigger than the first one: $\theta_2 > \theta_1$. The second constraint imposes that the incident beam and the reflected beams lie in the same plane. When the two beams are in the vertical plane, the incident angle μ is maximum and given by $\mu_{max} =$

$\theta_3 = 2(\theta_2 − \theta_1)$. Whatever the DCD settings, the beam illuminates the sample surface at point O. The distances between the crystals and the sample are also fixed so that the connected intervals $C_1C_2$, $C_1O$, and $C_2O$ form the triangle $OC_1C_2$ (see Fig. 2A). The incident angle $\mu$ is set by rotating the whole DCD setup by an angle $\rho$ around its main optical axis ($\rho$-axis), which is supposed to coincide with the primary beam. The angle between the beam after the second crystal and the horizontal plane of the sample is the beam grazing angle $\mu$ on the liquid sample surface, given by $\sin \mu = \sin \rho \sin \theta_3$. At $\rho=0$ the beam lies in the horizontal plane of the sample, thus $\mu=0$ (see Fig. 2B). The increase in angle $\rho > 0$ also increases $\mu > 0$ (see Fig. 2C), to finally reach the maximum value $\mu_{max} = \theta_3 = 2(\theta_2 − \theta_1)$ at $\rho=90°$ (see Fig. 2D).

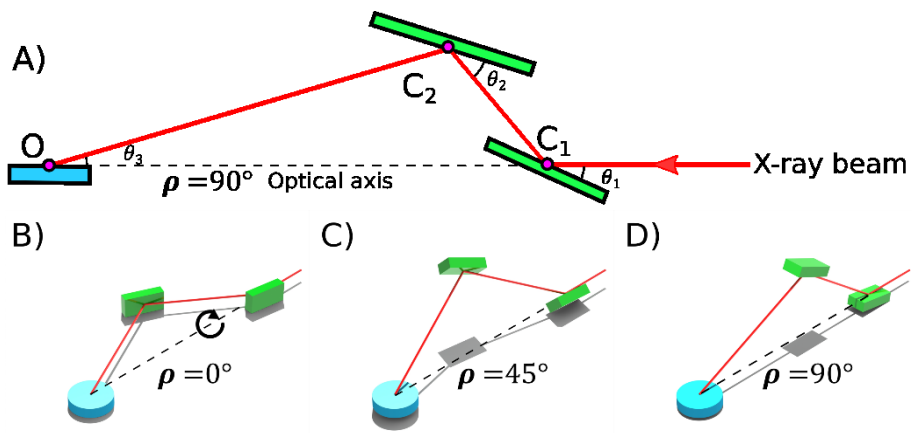

Figure 2 A) The geometrical sketch of the side view (vertical plane) on the DCD crystals assembly and sample at $\rho=90°$. B) The 3D drawing of the DCD configuration corresponding to µ=0° (i.e., $\rho=0°$), C) the intermediate situation when µ>0° (0° < $\rho$ < 90°), and D) at maximum µ_max ($\rho=90°$), situation corresponding to Fig. 1A. The arrow in B) shows the direction of the rotation of the crystals around the optical axis $\rho$.

Here we reach the crucial issue: any angular misfit between the primary incident beam and the optical axis $\rho$ will lead to a progressive loss of the Bragg condition, and thus of intensity, with varying $\rho$. Thus, this misfit has to be precisely measured and corrected before the XRR data collection, so that the DCD optical axis coincides with the primary beam. To overcome this issue, we calculate the angular drift analytically from the Bragg condition during the $\rho$ rotation around the optical axis with a non-zero misfit and apply a quantitative correction. The described situation is presented in Fig. 3. The blue line marks the DCD optical axis $\rho$, the X-ray beam propagates along the X-axis, and angles $\phi$ and $\omega$ are parasitic offsets of the DCD optical axis relative to the X-axis in the XY and XZ planes, respectively. The vector $\vec{n}$ is normal to the scattering plane of the first crystal, which initially, at $\rho=0$, makes an angle of $\pi/2 + \theta$ with the X-axis, i.e., is at the Bragg condition. In general, the vector $\vec{n}$ can be misaligned by a tilt angle $\tau$ relative to the XY plane. However, we assume that $\tau=0$, so that the initially diffracted beam propagates in the horizontal plane. The crystals of the DCD at ID10 are mounted on a manual stage to remove this parasitic tilt and to obtain the $\tau=0$ condition when the Bragg angle rotation axis is perpendicular to the horizontal plane. The angle variation between the vector $\vec{n}$ and the X-axis during rotation around the $\rho$-axis by angle $\rho$ can be easily obtained with the corresponding rotation matrix $R_\rho$:

$$R_\rho(\rho)=R_z(\phi)R_y(\omega)R_x(\rho)R_y(-\omega)R_z(-\phi) \qquad (1)$$

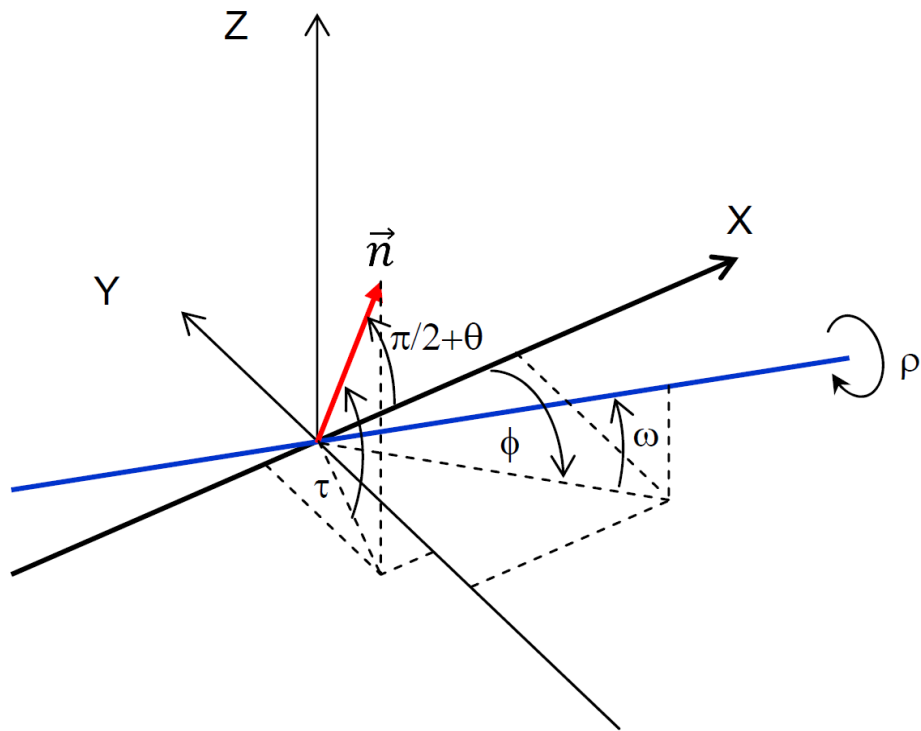

**Figure 3** Schematic sketch of the DCD geometry with a misfit. The black lines X, Y, and Z mark the laboratory coordinate system. The incident X-ray beam is along the X-axis. The blue line is the main DCD optical axis ($\rho$-axis), the red arrow marks the vector $\vec{n}$ normal to the scattering plane of the first crystal. The angles $\phi$ and $\omega$ are parasitic angular offsets of the $\rho$-axis from the X-axis (primary beam). The angle $\rho$ is the rotation angle of the whole DCD setup around its main optical axis. $\tau$ is the angle (assumed to be zero here) between the vector $\vec{n}$ and the XY plane.

Here $R_j, j \in \{x, y, x\}$ are rotation matrices around the respective coordinate axes. For an elementary rotation by an angle $\gamma$ around the corresponding axis, they are given by

$$R_x(\gamma) = \begin{pmatrix} 1 & 0 & 0 \\ 0 & \cos\gamma & \sin\gamma \\ 0 & -\sin\gamma & \cos\gamma \end{pmatrix}, R_y(\gamma) = \begin{pmatrix} \cos\gamma & 0 & -\sin\gamma \\ 0 & 1 & 0 \\ \sin\gamma & 0 & \cos\gamma \end{pmatrix}, R_z(\gamma) = \begin{pmatrix} \cos\gamma & \sin\gamma & 0 \\ -\sin\gamma & \cos\gamma & 0 \\ 0 & 0 & 1 \end{pmatrix}$$

(2)

In the described geometry, the X-ray beam orientation is expressed by the vector:

$$\vec{b} = \begin{pmatrix} 1 \\ 0 \\ 0 \end{pmatrix},$$

while the normal vector $\vec{n}$ to the scattering plane lying initially in the XY plane (i.e., $\rho=0$ and $\tau=0$) is expressed by the vector:

$$\vec{n_0} = \begin{pmatrix} -sin\theta \\ -cos\theta \\ 0 \end{pmatrix}.$$

Its coordinates are modified after rotation by the angle $\rho$ around the $\rho$-axis according to:

$$\vec{n}(\rho) = R_\rho(\rho)\vec{n_0} \tag{3}$$

We then derive the deviation angle $\varepsilon$ from the Bragg condition during a rotation $\rho$ around the $\rho$-axis from the equation:

$$\vec{n}(\rho) \cdot \vec{b} = -|\vec{n}(\rho)||\vec{b}|sin(\theta + \varepsilon) \tag{4}$$

The effect of the misfit between the $\rho$-axis and the X-ray beam is presented in Fig. 4, which shows the plot of the Bragg deviation angle $\varepsilon$ as a function of $\rho$, calculated using Equation (4) at $\theta = 4.5°$, $\phi = 0.002°$ and $\omega = 0.004°$.

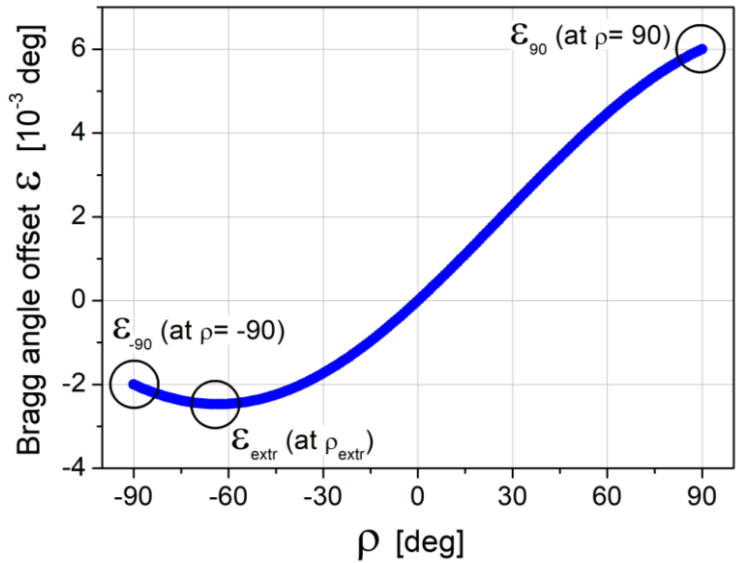

**Figure 4** The plot of $\varepsilon$ as a function of $\rho$, calculated using Equation (4) at $\theta = 4.5°$, $\phi =0.002°$ and $\omega = 0.004°$.

There are three crucial points on the graph: $\varepsilon_{+90}$ ($\varepsilon$ at $\rho= +90°$), $\varepsilon_{-90}$ ($\varepsilon$ at $\rho= -90°$), and $\varepsilon_{extr}(\rho_{extr})$ (position of the extremum). It is easy to show, using Equation (4) and the small-angle-approximation of trigonometric functions for small values of $\phi$, that:

$$\tan(\rho_{extr}) \cong \frac{\omega}{\phi} \tag{5}$$

$$\varepsilon_{+90} - \varepsilon_{-90} \cong -2\omega \cos\theta \tag{6}$$

Note that the angles $\phi$ and $\omega$ have to be expressed in radians in the equations. This result provides a straightforward procedure for the DCD alignment in order to make the $\rho$-axis to coincide with the incident X-ray beam. First, we measure the angle for the Bragg scattering on the first crystal at $\rho$= +90º and $\rho$= -90º. Following Equation (6), the difference between these two measured angles is a correction angle $\omega$. It is clear from Equation (5) that after rotation of the whole DCD assembly around the Y-axis by the correction angle $\omega$, the position of $\varepsilon_{extr}$ will be at $\rho$=0. So, for the final step of the DCD alignment, only two additional measurements of the $\varepsilon$ values at $\rho$= 0º and $\rho$= +90º are sufficient. The difference between these two values equals the sought correction angle $\phi$. After rotation of the DCD assembly around the Z-axis by this angle, the DCD alignment is completed. Routinely done DCD alignment is achieved with residual errors of $\omega \leq 0.5\ \mu rad$ and $\phi \leq 3.5\ \mu rad$.

The fine alignment of the DCD $\rho$-axis needs to guarantee that during the rotation its wobble remains significantly smaller than the angular acceptance (the so-called Darwin width) of the used crystals to preserve as accurately as possible the maximum intensity of the Bragg reflection for the whole operational energy range of the beamline. Figure 5 shows that for a pair of Ge(111) and Ge(220) crystals, a standard setup, the wobble value must be well below 15 $\mu$rad. With the fine optimization of the $\rho$-axis rotation stage, we usually achieve a wobble below 5 $\mu$rad (Fig. S1), i.e., far below the angular acceptance of the Ge(111) and Ge(220) pair of crystals at ID10, guaranteeing a well-tuned DCD scattering geometry.

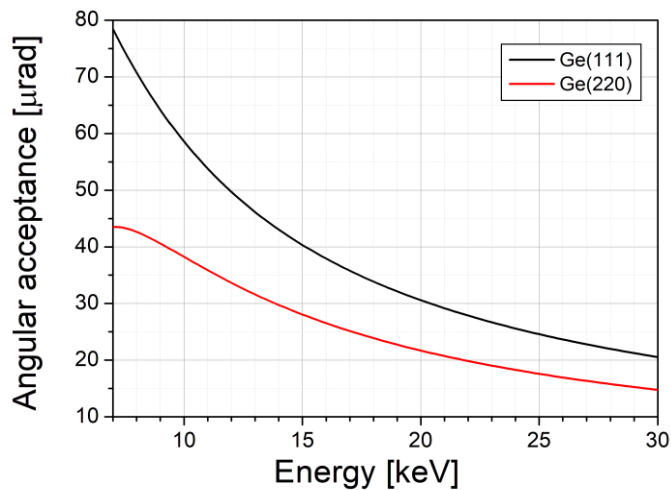

**Figure 5** Angular acceptance (Bragg peak Darwin width) of Ge(111) and (Ge(220) reflections versus X-ray energy.

With the very low wobble of the $\rho$-axis and our easy and fast method for the precise DCD alignment, the $q_z$ range tripling becomes straightforward. For this, we use three times higher-order reflections, namely, Ge(333) and Ge(660). Bragg angles of Ge(333) and Ge(660) at a given X-ray energy $E$ are respectively (almost) the same as for Ge(111) and Ge(220) at an X-ray energy of $E/3$, so the angle $\mu_{max} = \theta_3 = 2(\theta_2 - \theta_1)$ is the same for both energy configurations. However, due to the three times higher energy (or three times lower wavelength λ) in the case of Ge(333) and Ge(660), the $q_{max}$ is also three times higher ( $q_z^{max}(E) = 3 \cdot q_z^{max}(E/3)$ ). By using this approach, the $q_z^{max}$ range is increased from 2.5 Å$^{-1}$ to 7.5 Å$^{-1}$. The cost for the $q_z^{max}$ extension is less than three orders of magnitude loss of the X-ray beam intensity illuminating the sample. The main reason for this decrease in intensity is the weaker scattering and the narrower Darwin width of the higher-order Bragg peaks. However, this loss is not dramatic with the novel fourth-generation synchrotron sources, like the recently operating (since 2020) ESRF EBS (Raimondi, 2016). At ESRF beamline ID10, the measured X-ray beam photon flux is $10^{13}$ photons/sec before the DCD and about 7x10$^{10}$ photons/sec after the Ge(333) and Ge(660) reflection, at 22 keV. This beam intensity, with a cross-section of 26x10 $\mu m^2$, is sufficient to measure XRR up to $q_z^{max}$ on liquid metals.

Two types of XRR curves were recorded to verify the capability of extended range measurements at ID10. Fig.6 presents the XRR curves recorded *in situ* (at 1400 K) from bare liquid copper and liquid copper covered by a graphene monolayer. In Fig. 6A, the total scattering signal is plotted as a function of $q_z$. For the bare copper (blue curve), it is easy to distinguish the first-order peak, at $q_z = 3 \text{ Å}^{-1}$, and the broad second-order peak, with the maximum at $q_z = 5.5 \text{ Å}^{-1}$, with further signal decrease up to 7 Å$^{-1}$. These two broad peaks arise from sub-surface layering in the liquid (Magnussen *et al.*, 1995; Shpyrko *et al.*, 2005; Pershan & Schlossman, 2012) and the liquid bulk structure. In the case of the graphene layer (orange color), the curve is measured only up to $q_z < 4 \text{ Å}^{-1}$ because from 2.5 Å$^{-1}$ onwards the measured signal is dominated by the scattering from the bulk of liquid copper. The reconstructed specular rod intensity, after subtraction of the diffuse background, is plotted in Fig. 6B. Compared with bare copper, graphene-covered copper shows a pronounced minimum at $q_z = 0.8 \text{ Å}^{-1}$, in agreement with previous reports (Jankowski *et al.*, 2021). The specular reflection vanishes rapidly above $q_z > 1.7 \text{ Å}^{-1}$, as expected due to the surface roughness. However, the capability to measure up to very high $q_z$ values, where two structure peaks of liquid metals are accessible, allows the study of surface layering with better precision.

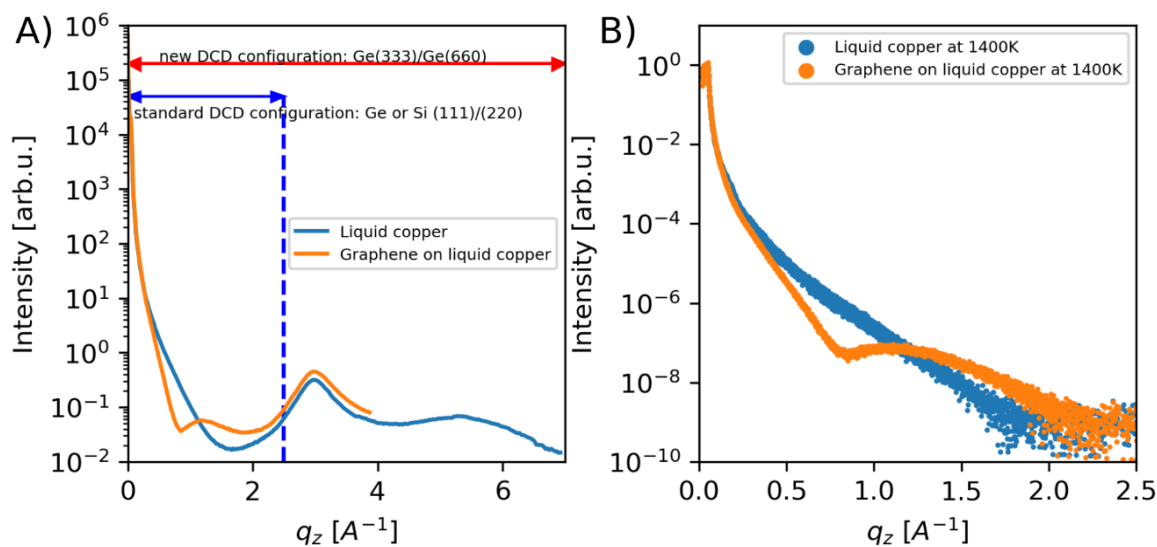

**Figure 6** A) Plot of the scattering intensity as a function of $q_z$ recorded from bare liquid copper (blue curve) and graphene-covered liquid copper (orange curve) at 1400 K. B) Specular rod, obtained after diffuse background subtraction, of bare liquid copper (red curve) and graphene-covered liquid copper (black curve) at 1400 K.

**Conclusions**

We have analytically described the misalignment correction of a double crystal deflecting system used to tilt the incident synchrotron X-ray beam with respect to the sample surface for grazing incidence scattering experiments on liquid surfaces. The proposed method is fast and straightforward, considering the complexity of the system and the demand for very high accuracy. In addition, we have developed a procedure that significantly extends the maximum range of momentum transfer perpendicular to the surface $q_z$, from ~2.5 Å$^{-1}$ to ~7 Å$^{-1}$. The new procedure is demonstrated for a bare and graphene-covered liquid copper surface. The recorded signal intensity is enhanced by the recent upgrade of ESRF to an EBS, allowing for more demanding measurements. The proposed method and the ESRF technical upgrade allow for new experiments with liquid metal surfaces and other systems. The measurements of out-of-plane crystallinity and order, i.e., Bragg peaks, Laue fringes, and strain effects, of materials like thin layers, nanoparticles, and quantum dots, supported on liquid surfaces, are now possible in the extended range of momentum transfer perpendicular to the surface.

**Acknowledgments**

The authors acknowledge the Huber Diffraktionstechnik GmbH & Co. KG company and Norman Huber in particular for the construction of the diffractometer-DCD assembly and the demanding work on the optimization of the DCD's rotation stage to reach our technical requirements. Furthermore, the authors appreciate the excellent work of ESRF engineers Muriel Magnin-Mattenet and Carol Clavel, and the ID10 beamline technician Karim Lhoste on all stages of the design, installation, and DCD tuning. On top of these, we thank the European Synchrotron Radiation Facility for providing synchrotron radiation facilities.

**Funding information**

The following funding is acknowledged: European Union's Horizon 2020 research and innovation program under Grant Agreement No. 951943 (DirectSepa)

**Supporting information**

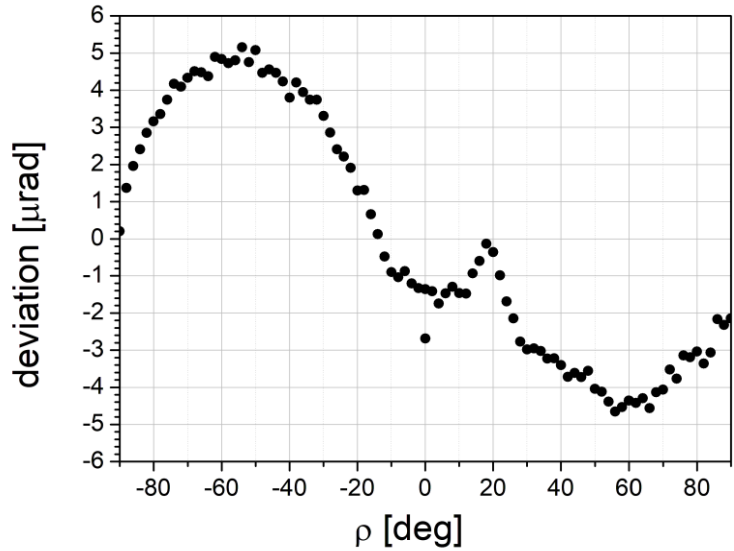

**Fig. S1 Experimentally measured deviation of the Ge(111) Bragg angle position versus ρ-axis rotation. Measurements were performed at an X-ray energy of 22 keV. ID10 DCD operates in the range between 0 and -90 degrees.**

**Note 1**

**About the maximum $q_z^{max}$ dependence on the X-ray energy for DCD setup**

The maximum of $q_z^{max}$ is:

$$q_z^{max} = 4\pi\lambda^{-1}\sin\mu_{max} . \qquad (1)$$

Here $\lambda$ is the X-ray wavelength and is $\mu_{max} = 2(\theta_2 - \theta_1)$, where $\theta_1$ and $\theta_2$ are the Bragg angles of the first and second crystals, respectively.

From the Bragg law $\theta_1 = arcsin\left(\frac{\lambda}{2d_1}\right)$ and $\theta_2 = arcsin\left(\frac{\lambda}{2d_2}\right)$, where $d_1$ and $d_2$ are lattice spacing for corresponding (h,k,l) indexes. Using this and formula (1), one can write:

$$\frac{\lambda q_z^{max}}{4\pi} = \sin\mu_{max}$$

$$arcsin\left(\frac{\lambda q_z^{max}}{4\pi}\right) = \mu_{max}$$

$$arcsin\left(\frac{\lambda q_z^{max}}{4\pi}\right) = 2\left(arcsin\left(\frac{\lambda}{2d_2}\right) - arcsin\left(\frac{\lambda}{2d_1}\right)\right) \qquad (2)$$

Using that $arcsin\, x - arcsin\, y = arcsin(x\sqrt{1-y^2} - y\sqrt{1-x^2})$ and omitting, for simplicity of further illustration, the constant factor on the left side of the formula (2), we can write

$$\frac{\lambda q_z^{max}}{4\pi} \propto \frac{\lambda}{2d_2}\sqrt{1-\left(\frac{\lambda}{2d_1}\right)^2} - \frac{\lambda}{2d_1}\sqrt{1-\left(\frac{\lambda}{2d_2}\right)^2} \qquad (3)$$

Formula 3 can be simplified to:

$$q_z^{max} \propto 2\pi \left( \frac{1}{d_2}\sqrt{1-\left(\frac{\lambda}{2d_1}\right)^2} - \frac{1}{d_1}\sqrt{1-\left(\frac{\lambda}{2d_2}\right)^2} \right) \quad (4)$$

With an increase of the X-ray energy, the wavelength decreases and asymptotically approaches 0 at infinite energy following relation $\lambda \propto E^{-1}$. Hence for small wavelengths, we can write:

$$q_z^{max} \propto 2\pi \left( \frac{1}{d_2} - \frac{1}{d_1} \right) \quad (5)$$

This shows that $q_z^{max}$ value does not depend much on the X-ray wavelength for sufficiently high energy. There is a more negligible energy dependence for lower energies. However, this effect is insignificant for the energy range at which the ESRF beamline ID10 operates, i.e., 7-30 keV. Figure 2 shows the $q_z^{max}$ value as a function of energy, calculated for two pairs of crystal reflection Ge(111)/Ge(220) (black curve and the left axis) and Ge(333)/Ge(660) (red curve and the right axis)

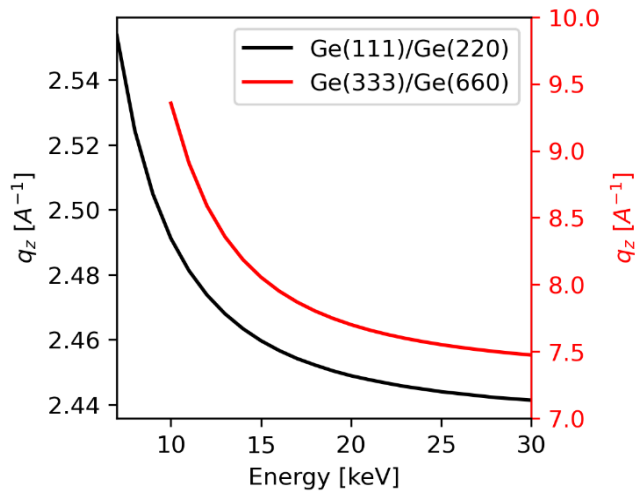

**Fig. S2 The value of the $q_z^{max}$ plotted as a function of X-ray beam energy. The black plot and axis correspond to values calculated for the pair of Ge(111)/Ge(220) reflections. The red plot and axis correspond to values calculated for the pair of Ge(333)/Ge(660) reflections.**